\documentclass[aps,pra,twocolumn,showpacs,floatfix,superscriptaddress]{revtex4}
\usepackage{graphicx}
\usepackage{soul}
\usepackage[usenames,dvipsnames]{color}
\usepackage{amsmath}
\usepackage{multirow}
\newcommand{\eq}[1]{Eq.~(\ref{#1})}
\newcommand{\fig}[1]{Fig.~\ref{#1}}
\newcommand{\tab}[1]{Table~\ref{#1}}
\newcommand{\sect}[1]{Section~\ref{#1}}
\newcommand{\be}[1]{\begin{equation}\label{#1}}
\newcommand{\ee}{\end{equation}}

\newcommand{\blue}[1]{\textcolor{blue}{#1}}

\def\li#1{\overline{#1}}
\def\lin{{\overline{n}}}

{\catcode`\|=\active 
  \gdef\Braket#1{\left<\mathcode`\|"8000\let|\bravert {#1}\right>}}
\def\bravert{\egroup\,\vrule\,\bgroup}

\date{\today}
\begin{document}
\title{Electronic excitation by short X-ray pulses:\\
From quantum beats to wave packet revivals}
\author{Paula Rivi{\`e}re}
\affiliation{Max-Planck Institute for the Physics of Complex Systems\\
  N{\"o}thnitzer Str. 38, 01187 Dresden, Germany.}
\affiliation{Departamento de Qu{\'i}mica, Universidad Aut{\'o}noma de Madrid, 28049 Madrid, Spain.}

  \author{Shahid Iqbal}
\affiliation{Max-Planck Institute for the Physics of Complex Systems\\
  N{\"o}thnitzer Str. 38, 01187 Dresden, Germany.}

\author{Jan M Rost}
\affiliation{Max-Planck Institute for the Physics of Complex Systems\\
  N{\"o}thnitzer Str. 38, 01187 Dresden, Germany.}
  \affiliation{Advanced Study Group at the Center for Free Electron Laser Science, DESY, Hamburg, Germany}
\date{\today}
\begin{abstract}
\noindent
We propose a simple way to determine the periodicities of wave packets in quantum systems 
directly from the energy differences of the states involved.
The resulting classical periods and revival times
are more accurate than those obtained with the traditional expansion of the energies about the central quantum number $\overline{n}$,
especially when  $\overline{n}$ is low. The latter type of wave packet motion occurs upon excitation of highly charged ions with  short XUV or X-ray pulses. Moreover, we formulate the wave packet dynamics in  such a form that it directly reveals the origin of phase shifts in the maxima of the autocorrelation function. This phenomenon, so far poorly understood since it is not prominent in the high $\overline{n}$ regime, becomes a dominant feature in low $\overline{n}$ wave packet dynamics.
\end{abstract}
\maketitle
\section{Introduction}
 A wave packet (WP) comprising many energy eigenstates of a system spreads during its time evolution,
only to reverse the spreading and reshape after a certain time called the revival time.
Originally formulated in the context of highly excited electrons in atoms \cite{Alber1986} following the first
observation of wave packet collapse and revival \cite{Parker1986},
the phenomenon has been identified in a large variety of physical systems, such as in Gaussian WPs in quantum boxes \cite{Grossmann1997}, the evolution of
rovibrational nuclear WPs \cite{Ergler2006,De2010} (for a review see \cite{Calvert2010}), even at an attosecond time scale \cite{Katsuki2010}, revivals in a coherent photon field \cite{Keeling2008}, or  recently, the propagation of WPs in graphene under magnetic fields \cite{Krueckl2009}.

Quantum revivals have also been widely studied over the years in a more mathematical context,
with emphasis on obtaining accurate analytical expressions  \cite{Leichtle1996,Aronstein2005}, for a review see \cite{Robinett2004}.

The multiple interferences of the wave packet components lead also to the so-called fractional revivals at divisors
of the revival time, which have been predicted \cite{Averbukh1989,Veilande2007} and observed in Rydberg wave packets
\cite{Yeazell1991} or photon bouncing balls \cite{Valle2009}.
Fractional revivals can retrieve information about the system even if it decays before the first revival \cite{Ghosh2007}.
They can be used for mapping the quantum phase of a molecular nuclear wave-packet 
in two-dimensional spectroscopy \cite{Schubert2010}, 
or even to factorize prime numbers \cite{Merkel2006}. 
It has also been shown that information entropy in position and momentum spaces can reveal the existence of fractional revivals \cite{Romera2007}.


The study of quantum revivals in atoms has  traditionally focused on the excitation of Rydberg WPs. 
The latter contain typically of the order of 10  highly excited states when they are centered on a principal quantum number $30 < \overline{n} <85$ and excited by laser pulses with a duration of a few picoseconds \cite{Alber1986}. Under these circumstances the relative difference between the energies of the states is small, and the standard approach of representing the energies of the contributing states by Taylor expansion about the central energy $E_{\overline n}$ provides good results.

However, new FEL or high harmonic-based light sources are able to deliver short pulses of 100 atomic units (2.4 fs) duration and less in the VUV to X-ray regime \cite{Ayvazyan2005}. 
In particular the X-ray pulses will typically lead to a multi-electron wave packet of the valence shell of an atom by 
inner shell ionization. In order for the photo electron to form a bound wave packet while simultaneously the core electrons remain in their ground state, such wave packets must be generated in highly charged ions, as produced,  e.g., in state-of-the-art electron-cyclotron ion sources (ECR) \cite{Knoop2005} or EBITs \cite{Fischer2002}. 
The high ionic charge implies large electronic energy spacings  so that only a few levels contribute to the wave packet, despite the large energy width due to the short pulse duration. Hence, the dynamics lays in between quantum beats of a few well defined energy levels \cite{Averbukh2010} and the traditional regime of Rydberg wave packets as discussed before.  

\begin{figure}
\includegraphics[width=0.23\textwidth]{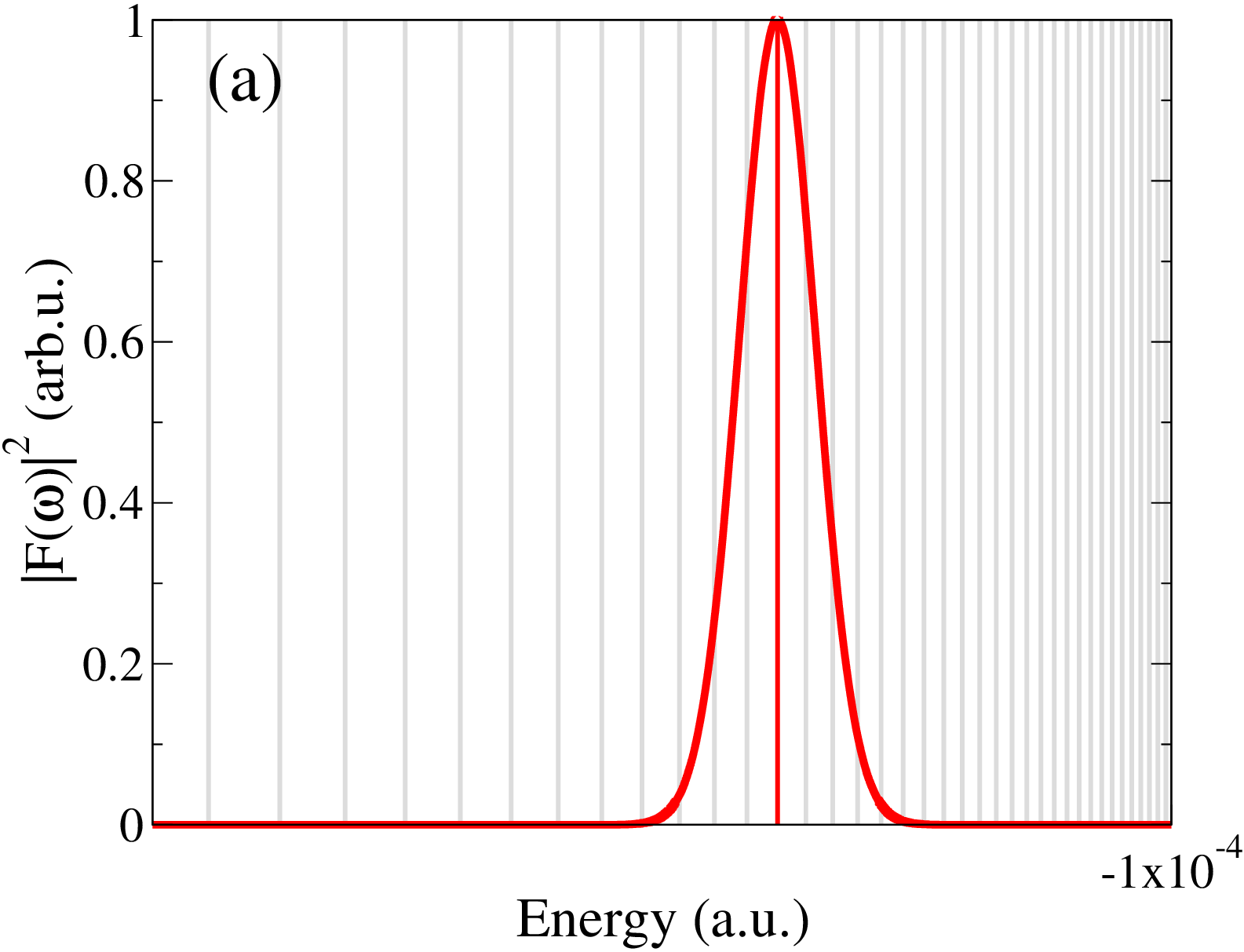} 
\includegraphics[width=0.22\textwidth]{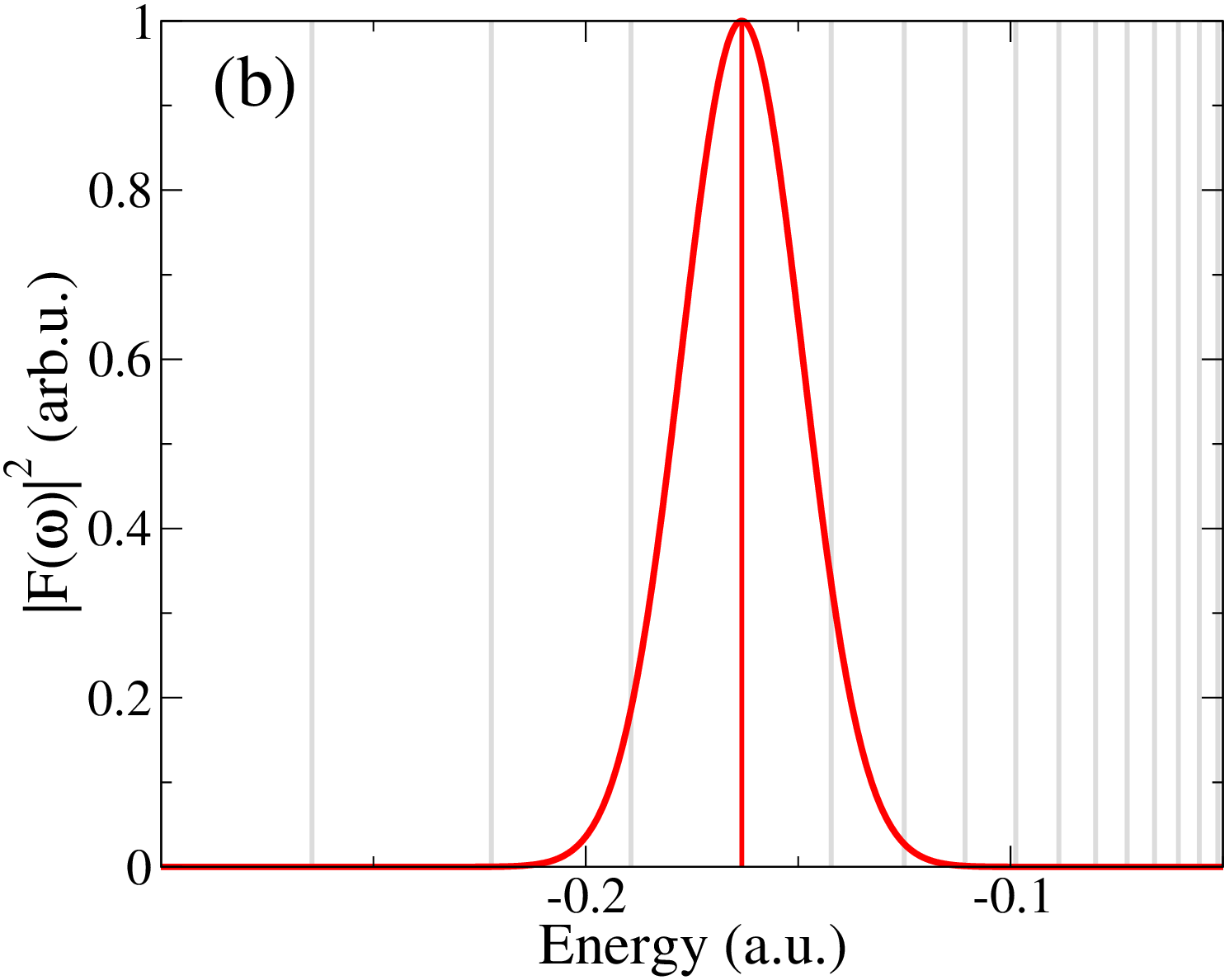} 
\caption{(Color online) Field intensities in the frequency regime, $|F(\omega)|^2$, and energy levels
for two regimes: (a) high $\li{n}$,  long pulses, for the particular
case of $Z=1$, $\li{n}=45$, FWHM=2 ps, 
(b) low $\li{n}$,  short pulses, for the case of $Z=8$, $\li{n}=14$, FWHM=2 fs. The red vertical lines
correspond to $\li{n}$ in each case. For the number of energy levels participating in the wave packet with an amplitude of more than $10^{-2}$, see table 1.} 
\label{lowhigh2}
\end{figure}

A good estimate of the dynamical range of a system, which goes from quantum beats to wave packet revivals, is the number of states in the WP, $k$. One can estimate this number by relating the energy spread of the laser pulse
to the energy spread in the excited atom, which for a WP centered in state $\lin$ is
 $\Delta E  \simeq k\, dE_{n}/dn|_{\lin}$.
An electric field of Gaussian shape 
$f(t)=f_0\exp^{-2\ln2 t^2/T^2}$,
where $T$ is the FWHM of the pulse and $f_0$ its amplitude,  will populate a wave packet with amplitude 
\begin{equation}\label{Gaussian}
F_n=f_0\exp^{-(E_{n}-E_{\lin})^2 T^2/(8\log{2})}
\end{equation}
for each energy level $E_n$.
Including all those states with probability $\geq c$ implies an energy spread of $\Delta E=C/T$,
where $C = 4(-\ln 2\ln c)^{1/2}$.
For a hydrogenic ion of charge $Z$ and spectrum $E_{n} = -Z^2/(2n^2)$, equalling both values of $\Delta E$ gives
\begin{equation}
\label{k-estimate}
k = C \overline n^3/(Z^2 T).
\end{equation}
A few quantitative examples of the number of states $k$ as a function of $\lin$ and $T$ for different atomic numbers is shown in Table 1. Two cases of low and high $\lin$ are shown in \fig{lowhigh2}.

\begin{table}
\caption{\label{table 1} Number of states $k$ contributing to a wave packet  with  up to an amplitude  weight of $ c =10^{-2}$  when excited to a central state $\lin$ by a pulse of duration T (FWHM). We assume an hydrogenic ion with energy spectrum $E_{n} =   -Z^2/(2n^2)$. Parameters given are used in the figures of this paper as indicated.}
\begin{ruledtabular}
\begin{tabular}{rrrlc}
k & Z &$\lin$  &  T [fs] & Figs.\\\hline
13 &1 &  85 &     $8000$ & 1\\
8 &1 &  45 &     $2000$ & 4,5\\
4 &8 &  14 &     $2$ & 4,5\\
4 &12 &  7 &     $0.1$ & 6\\
5 &8 &  9 &     $0.4$ & 8\\
\end{tabular}
\end{ruledtabular}
\end{table}

Our goal is to treat the low $\lin$ regime of electron excitation from the perspective of wave packet revivals. However,
due to the large energy spacing  the standard description of the WPs with a harmonic approximation about $E_\lin$ becomes rather inaccurate.
In the following we will show that one can  replace the derivatives of the energies
by exact differences of energy levels to obtain a description which is valid in the
new regime of short, high frequency pulses. This approach is conceptually very simple and it is universally valid, also in the traditional Rydberg regime. We can derive 
 accurate values for the revival times in all situations. Moreover,
the origin of the shift in the position of the maxima around the revival time, previously noticed \cite{Averbukh1989,Yeazell1990} but not well described, emerges naturally.

The structure of the paper is as follows: in section \ref{sec:at}, a form of the autocorrelation function is derived which allows one to explicitly read off the relevant periods.
We also derive and explain why for the low excitation regime finite energy differences should be used
to define these periods instead of energy derivatives.
The periods,
classical,  revival and fractional revival times, as well as a  feature which is important for low excitation wave packets, namely the time where the extrema flip their sign, are explained in section \ref{sec:period} with examples from the high and intermediate excitation domain.
 Section \ref{sec:low} points out the characteristic features of low excitation wave packets  ending with
 a hydrogenic ion excited by a 100 attosecond X-ray (1.92 keV) pulse. Finally, for multi-electron atoms,
the effect of the quantum defect is discussed in \ref{sec:qd}, and
 conclusions are drawn in \ref{sec:conc}.

\section{Autocorrelation function}\label{sec:at}
Without loss of generality but to be somewhat specific we
 consider  a system which absorbs one photon  from a laser pulse of length $T$ in a transition from an initial state $|\alpha\rangle$ to a final state  $|\lin\rangle$ which, for simplicity, we assume to be an eigenstate of the system.  
Due to the finite length of the pulse, this process creates a wave packet  which  can be expressed in 
 first order perturbation theory 
 after the pulse as 
\begin{eqnarray}
|\Psi(t)\rangle
&=&i\sum_n |n\rangle \exp[-iE_n t] M_{\alpha n}F_n,
\label{psii}
\end{eqnarray}
where $F_n$ are the amplitudes defined in \eq{Gaussian}. 
The dipole matrix elements $M_{\alpha n}=\langle n|z|\Psi_\alpha(0)\rangle$ give the strength of photon coupling to the different eigenmodes. Atomic units are used throughout the paper unless otherwise stated.

The  modulus square of the autocorrelation function  
\begin{equation}
A(t)=\left|\langle \alpha| \Psi(t)\rangle \right|^2= \sum_{n,m} |c_n|^2|c_m|^2  \exp[-i(E_n-E_{m}) t]
\label{atorig}
\end{equation}
with coefficients $|c_n|^2=|M_{\alpha n}|^2 |F_n|^2$ is 
formally equivalent to $P_{\alpha\rightarrow f}(t)$, the probability of going
from a state $|\alpha\rangle$ to a state  $|f\rangle$ in a pump-probe experiment involving two identical pulses with a time delay $t$. Therefore, \eq{atorig} can be  measured \cite{Alber1986}. 

One of the salient features of \eq{atorig} is that this time-dependent function shows revivals and partial
revivals of the initial wave packet. 
Clearly not all of the infinitely many terms in \eq{atorig} do contribute equally, since the pulse excitation covers only a finite energy interval of the spectrum $\{E_n\}$, see \eq{Gaussian}.  Hence, our goal is to rewrite
the double sum such that a systematic truncation to the most important terms is facilitated,
yet retaining in the most accurate manner the different periodicities which give rise to the revival phenomena.

To reorder the contributions in the sum, we will make use of an expansion of the eigenenergies $E_{n}$ around the mean
excitation $E_{\lin}$, which is well justified as long as the pulse excites  a region of high density of states,
so that the relative level spacing  $(dE/dn|_{\lin})/E_{\lin}\ll 1$. 

In a second step we will acknowledge the fact that for high frequency pulses the spectrum of highly charged ions is excited in a region of much lower density of states than in the traditional Rydberg regime.

\subsection{Regrouping the sum}\label{sec:regroup}

First we perform a Taylor expansion of the energy around the mean excitation energy $E_{\lin}$, and
split the expansion into two terms $\delta^{\pm}_{n}$ with even/odd powers of $n$, 
so that the energy differences become (see appendix \ref{app:terms})
\begin{subequations}
\begin{equation}\label{expansion}
E_{\lin+n}-E_{\lin} \equiv \delta_n = \delta^+_n+\delta^-_n
\end{equation}
where
\begin{equation}
\delta^+_n=2\pi\sum_{j=1}^{\infty} \frac{n^{2j}}{\bar{T}_{2j}},\quad 
\delta^-_n=2\pi\sum_{j=1}^{\infty}\frac{n^{2j-1}}{\bar{T}_{2j-1}}
\end{equation}
and
\begin{equation}\label{invT}
\frac{1}{\bar{T}_{j}}=\frac{1}{2\pi}\frac{1}{j!} \left| \frac{d^{j}E_{n+\lin}}{dn^{j}}\right|_{n=0}\,.
\end{equation}
\end{subequations}

For the regrouping  we assume that the dipole matrix elements $M_{\alpha\nu}$ do not vary much
in the relevant terms of \eq{atorig} and are therefore irrelevant for the summation. 
Moreover, the diagonal parts $m=n$ will only contribute a non-oscillating background, so that the time-dependent part of \eq{atorig} simplifies to
\be{ats}
a(t) = \sum_{n\ne m} W_{mn}\exp[-i(\delta_m-\delta_n)t]\,,
\ee
with weights
\be{weight}
W_{mn}=\exp[-(\delta_{m}^{2}+\delta_{n}^{2})/\sigma^{2}]
\ee
decreasing rapidly with increasing  distance $\delta_{|n|}$ from the central energy.
Here $\sigma=\sqrt{4\log{2}}/T$ is the pulse width in frequency domain, see \eq{Gaussian}.
The idea is now to arrange the double sum \eq{ats} into groups of terms which have similar Gaussian weight
$W_{nm}$,  where we order the weight according to $\delta_{m}\propto |m|$.
 From a basic element  $W_{nm}$ a quadruple of elements is generated by the two symmetry operations and their concatenation: exchange of the indices $(n,m)\to(m,n)$ (which amounts to complex conjugation),  and   point inversion $(n,m)\to (-n,-m)$. Note that $m,n$ can be negative due to the shift of the indices by the value $\lin$.  Hence we may rewrite \eq{ats}
as
\be{ats1}
\frac{a(t)}{2}  = \sum_{0\leq n<m}W_{mn}\cos [(\delta_{m}-\delta_{n})t]
+W_{-m-n}\cos[ (\delta_{-m}-\delta_{-n})t]\,.
\ee
Making use of the symmetry properties of the $\delta_{n}$ from \eq{expansion}, we get (see appendix \ref{app:final})
\begin{align}\label{ats2}
\frac{a(t)}{4} =\sum_{0\leq n<m}&e^{-\frac{w_{mn}}{\sigma^{2}}}\left\{
 C_{mn}\cos[(\delta^{+}_{m}-\delta^{+}_{n})t]\cos[(\delta^{-}_{m}-\delta^{-}_{n})t] \right.
  \nonumber\\ +& \left.S_{mn}\sin[(\delta^{+}_{m}- \delta^{+}_{n})t]\sin[(\delta^{-}_{m}-\delta^{-}_{n})t]\right\} \,,
\end{align}
where
\begin{subequations}
\begin{equation}\label{ats2a}
w_{mn}=(\delta^{+}_{m})^{2} +(\delta^{-}_{m})^{2}+(\delta^{+}_{n})^{2}+(\delta^{-}_{n})^{2} 
\end{equation}
accounts for the even and odd exponents of the neighboring levels, and
\begin{eqnarray}
C_{mn}&=&\sum_{j=0}^{\infty}\frac{(\delta^{+}_{m} \delta^{-}_{m}+\delta^{+}_{n}\delta^{-}_{n})^{2j}}{(2j)!\,\sigma^{4j}}2^{2j}\nonumber\\
S_{mn}&=&\sum_{j=0}^{\infty}\frac{(\delta^{+}_{m}\delta^{-}_{m}+\delta^{+}_{n}\delta^{-}_{n})^{2j+1}}{(2j+1)!\,\sigma^{4j+2}}2^{2j+1}\,
\end{eqnarray}
\end{subequations}
are the coefficients of the cosine and sine terms in \eq{ats2}, respectively.

\subsection{Expansion to lowest order around the mean energy}\label{sec:firstorder}

The final form of the autocorrelation function \eq{ats2} is still an exact version of \eq{ats}. While of little practical use, it reveals the structure quite well, which contains products of two trigonometric functions. Their meaning
becomes obvious if we truncate the Taylor expansion  \eq{expansion} around $E_{\lin}$ to lowest order,
\be{truncation}
\delta^{+}_{m}\simeq \frac{2\pi m^2}{\bar{T}_2},\quad \delta^{-}_{m}\simeq \frac{2\pi m}{\bar{T}_1}.
\ee
 Keeping only products of derivatives to second order, i.e., in times $T_{1}^{-1}, T_{2}^{-1}$ and $T_{1}^{-2}$, reduces  \eq{ats2} to (see appendix \ref{app:lowest}) 
 \begin{align}\label{ats-trunc}
 \frac{a_{2}(t)}{4} \quad \simeq& \sum_{0\leq n<m}e^{-\frac{4\pi^{2}\sigma_{T}^{2}}{T_{1}^{2}}(m^{2}+n^{2})}\\\nonumber
 &\times \cos\left[2\pi(m^{2}-n^{2})\frac{t}{\bar{T}_{2}}\right]\cos\left[2\pi(m-n)\frac{t}{\bar{T}_{1}}\right]\,.
 \end{align}
 From this representation one can easily read off the standard essential periodicities in the autocorrelation function: the classical and revival periods.
 One can also directly see how periodic maxima of $a(t)$ turn into minima for certain time intervals,
 as we will show in the next section.  
However, as we will also see, a further simplification using 
 the two energy values $E_{\lin \pm1}$ next to the mean energy $E_{\lin}$ can be made to construct
the autocorrelation function $a_{2}(t)$. 
Notice that this is not an additional approximation:
the phase in the cosine in \eq{ats1}, which is exact, is $(\delta_m-\delta_n)t=(E_{\lin+m}-E_{\lin+n})t$, so the
discrete energy differences are the ones which determine the periodicity of the system.


\subsection{Expansion to lowest order using finite differences in the energy eigenstates}\label{sec:finite}
The autocorrelation function in \eq{ats2} contains energy derivatives (see \eq{invT}). While they are
usually taken as continuous derivatives, we will consider finite differences
\begin{eqnarray}\label{finite}
E'_{\lin} &\equiv& \frac{(E_{\lin+1}-E_{\lin -1})}{2} 
= \frac{(\delta_{1}-\delta_{-1})}{2}\\\nonumber
\frac{E''_{\lin}}{2}&\equiv& \frac{(E_{\lin +1}  - 2E_{\lin}  + E_{\lin-1})}{2} 
= \frac{(\delta_{1}+\delta_{-1})}{2}
\end{eqnarray}
The motivation for this comes from the fact that 
for the X-ray excitation regime, the few states which contribute to the
autocorrelation function (see table 1) are not necessarily narrowly spaced in energy, so that a Taylor expansion may not be very accurate. 

On the other hand, using finite differences implies that one includes the most important terms in the autocorrelation function with \textit{exact} phases. 
The most important pair is $(m,n)=(1,0)$, since it gives rise to the terms with highest weights 
in the autocorrelation function, see \eq{weight}. For this pair of indices, the phases in \eq{ats-trunc} become
$E''_{\lin}/2$ and $E'_{\lin}$ (\eq{finite}). Therefore, 
these indices contribute to the autocorrelation function with the phases
\begin{align}\label{finitepair}
&2\cos\left[\frac{\delta_{1}-\delta_{-1}}2t\right]\cos\left[\frac{\delta_{1}+\delta_{-1}}2t\right]= \nonumber \\
&\cos\left[(\delta_{1}-\delta_0)t\right]+\cos\left[(\delta_{-1}-\delta_0)t\right],
\end{align} 
which are the phases of the two lowest order terms in \eq{ats1} (note that $\delta_0=0$). The amplitude is  slightly off
due to truncating $C_{mn}$ in \eq{ats2} to the $j=0$ contribution, see appendix \ref{app:lowest}. However, this does not affect the periodicities of $a(t)$, which we will discuss next.

\subsection{Difference of periodicities defined by continuous derivatives and finite differences}
\begin{figure}
\includegraphics[width=0.4\textwidth]{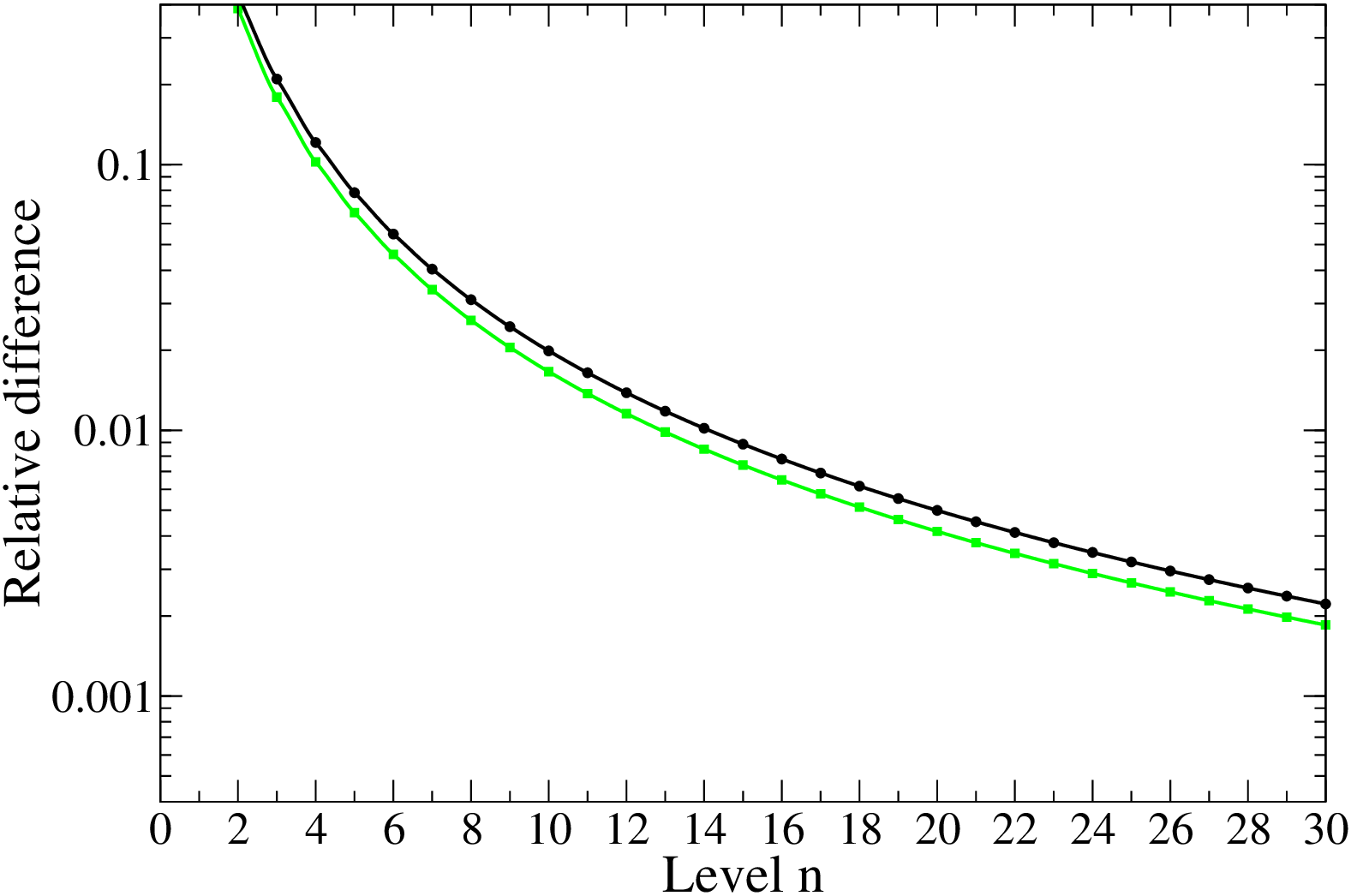} 
\caption{(Color online) Relative differences between the standard approach (continuous derivatives) and our method, in logarithmic scale.
Black line: classical period (\eq{difftc}). Green line: revival time (\eq{difftr}).}
\label{reldif}
\end{figure}

A natural question is how much the periods $\bar{T}_{i}$ defined by the continuous derivatives in 
\eq{expansion} differ from the periods $T_{i}$ defined by finite differences. This can be best illustrated
with an example, for which we take a hydrogenic Rydberg electron with spectrum
 $E_{n}=-Z^2/(2n^2)$ excited to $E_{\lin}$. The standard definition for the classical
period, understood as the orbiting period of a classical electron, is derived from the energy derivative
\begin{equation}
\bar{T}_c \equiv \bar{T}_{1}=  \frac{2\pi}{E'_{\li{n}}} = \frac{2\pi \li{n}^3}{Z^2}, 
\label{tccontRyd}
\end{equation}
while from finite differences we obtain
\begin{equation}
T_c \equiv T_{1} = \frac{4\pi}{E_{\lin+1}-E_{\lin -1}} = \frac{2\pi (\li{n}^2-1)^2}{\li{n}Z^2}.
\label{TcRyd}
\end{equation}
The relative difference between both expressions is
\begin{equation}
\frac{\bar{T}_c-T_c}{\bar{T}_c} = \frac{2}{\li{n}^2}-\frac{1}{\lin^4}, 
\label{difftc}
\end{equation}
which is noticeable for lower values of $\li{n}$: it is around
$1\%$ for $\li{n}=14$ (with a relative level spacing $\Delta E_{\lin}/E_{\lin}=2/\lin =0.3$), but less than $0.1\%$ for $\li{n}=45$ (with $\Delta E_{\lin}/E_{\lin} =0.07$). 
Similarly the revival times are
\begin{eqnarray}
\bar{T}_r  \equiv \bar{T}_{2} &=& \frac{4\pi \li{n}^4}{3Z^2},\\\nonumber 
T_r \equiv T_{2} &=& \frac{4\pi \li{n}^2(\li{n}^2-1)^2}{Z^2 (3\li{n}^2-1)}, 
\end{eqnarray} 
with a relative difference of
\begin{equation}
\frac{\bar{T}_r-T_r}{\bar{T}_r} = \frac{3}{\li{n}^2} - \frac{4}{3\li{n}^2-1}\,.
\label{difftr}
\end{equation}
This amounts to $\sim 0.85\%$ difference  for $\li{n}=14$, and only $0.08\%$ for $\li{n}=45$.
Hence, both expressions merge for high quantum numbers, while for low quantum numbers with a larger spacing between energy levels a finite difference is a poor approximation to a derivative, see the sketch in \fig{lowhigh2}.
In our case, the energy differences are the ``true" ones, so continuous derivatives are an approximation, which fails for low quantum numbers, as illustrated in \fig{reldif}.
For this reason, we will use in the following the  $T_{i}$ derived from
finite energy differences instead of the $\bar{T}_{i}$.

\section{Four essential periodicities of the autocorrelation function}
\label{sec:period}
\subsection{Classical period and revival time}
The autocorrelation function \eq{ats2} or its truncation \eq{ats-trunc} reveal directly
the classical period $T_{c}\equiv T_{1}$, for whose multiples the second cosine factor in each term becomes maximal.
These are the two fundamental time scales emerging from the first and second order energy differences
of energy levels next to the centrally excited one, as shown above.
 
\begin{figure}
\includegraphics[width=0.4\textwidth]{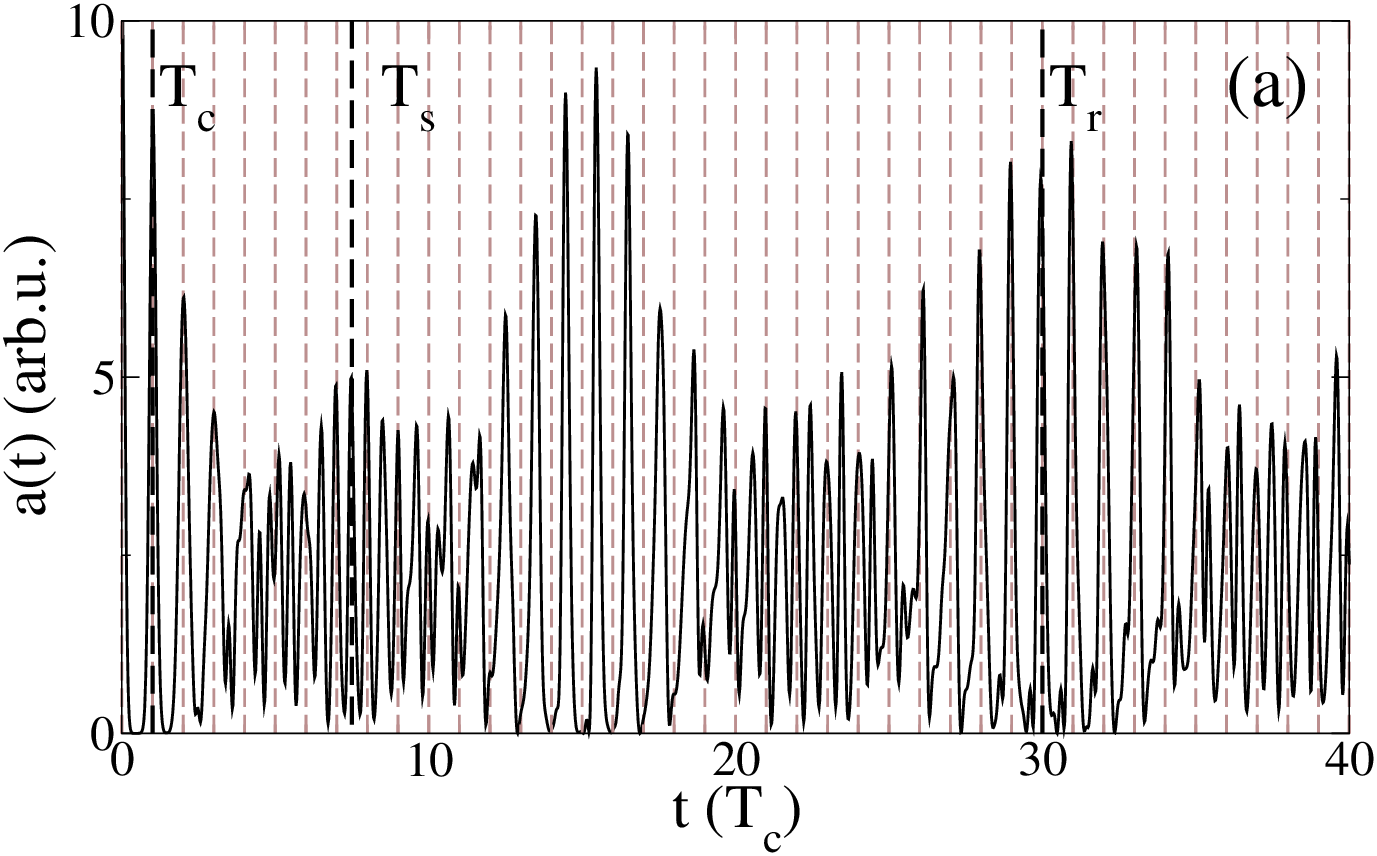} \\ 
\includegraphics[width=0.4\textwidth]{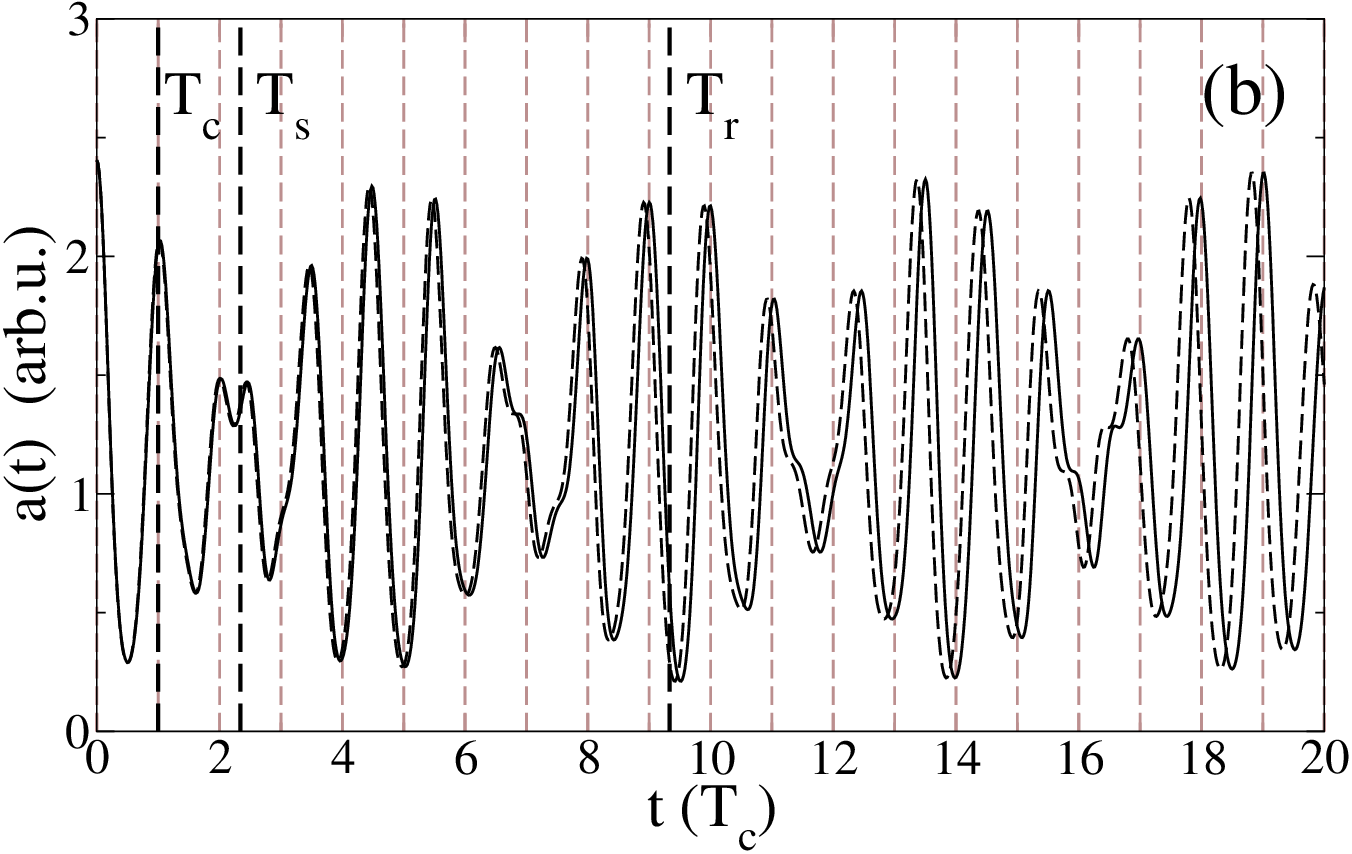} 
\caption{Autocorrelation function for (a) $Z=1$, $\li{n}=45$,  FWHM=2 ps, 
and (b) $Z=8$, $\li{n}=14$,  FWHM=2 fs. Full lines: versus $T_c$. Dashed lines:
versus $\bar{T}_c$. In (a) both are indistinguishable. The classical period $T_c$, revival time $T_r$, and the
time for sign change $T_s$ are shown with vertical lines.}
\label{autoRyd}
\end{figure}

  A representative autocorrelation function for a radial Rydberg wave packet which exhibits these well known features is shown in \fig{autoRyd}, for the high- and low-$\lin$ regimes. The classical and revival times are indicated.  
  Note that results for  periods $T_{i}$ and $\bar{T}_{i}$ coincide to the accuracy of the figure in the high-$\lin$ case (\fig{autoRyd}a), while a discrepancy is visible in the low-$\lin$ case of \fig{autoRyd}b.

 \subsection{The time $T_{s}$ of sign change}
 \label{sec:signchange}
 
 At first glance, the wave packet in \fig{autoRyd}a shows a revival at $T_{r}/2$,
 but a closer inspection reveals that this is not an exact copy of the wave packet, 
since its maxima are shifted by half a period $T_{c}$ (in between the vertical dashed lines).
 The revival at $T_{r}$ is complete, with maxima close to full periods of $T_{c}$. Although noticed before \cite{Yeazell1990,Averbukh1989}, there has been no formulation of the autocorrelation function which would reveal the shift directly.  This is achieved in  \eq{ats} with the products of two trigonometric functions and the cosine product dominating as revealed by \eq{ats2}. The first cosine function with its slow
 period  $T_{r}$ can be interpreted as an envelope to the rapidly oscillating second one with period $T_{c}$. 
Hence, we identify a time for sign change $T_{s}\equiv T_{r}/4$ at which the sign of the envelope changes, turning the maxima of the second cosine function into minima. This is the origin of the apparent shift in the position of the maxima.
 
 \subsection{Partial revivals}
 
\begin{figure}
\includegraphics[width=0.4\textwidth]{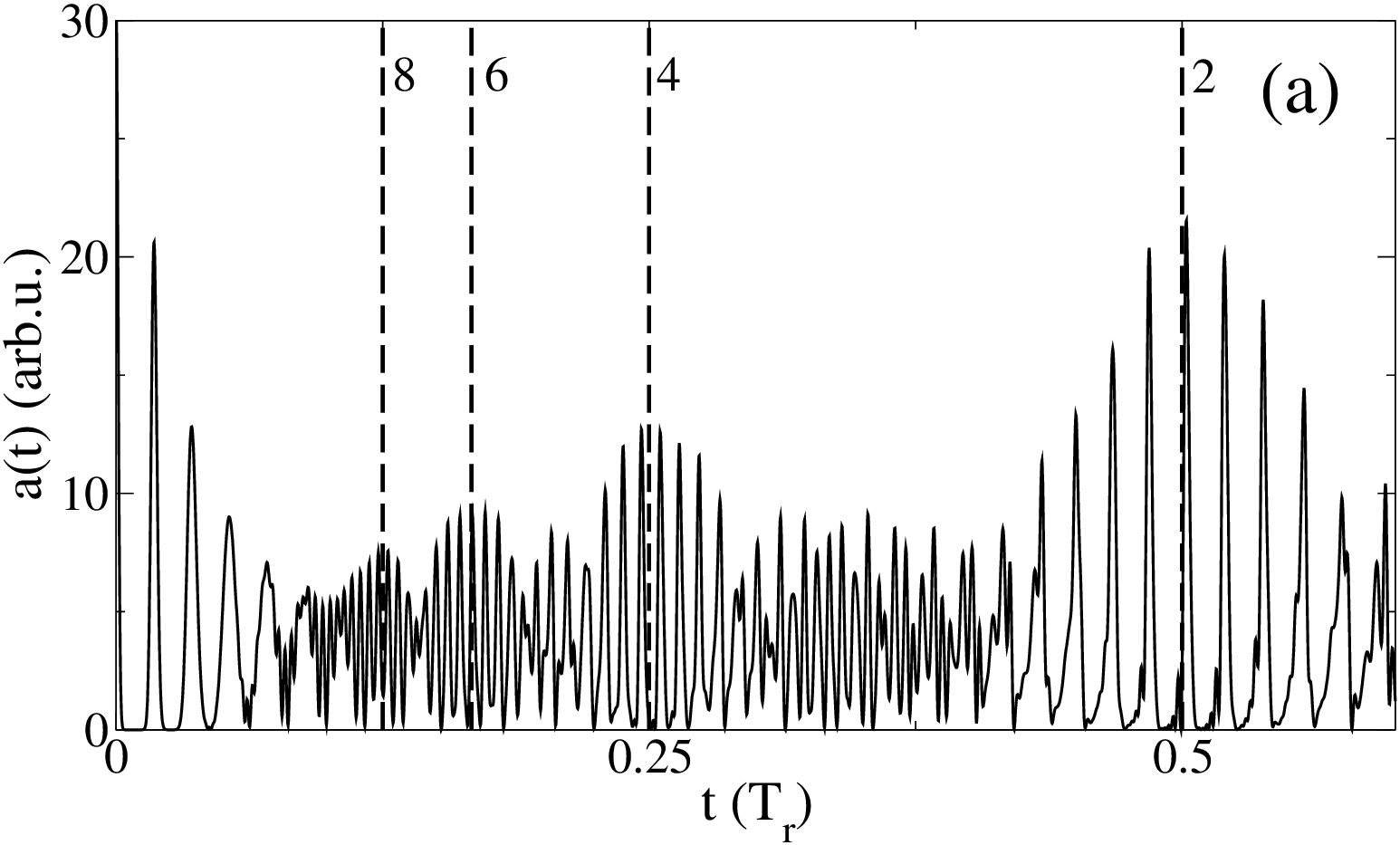} 
\caption{Autocorrelation function for a Rydberg wave packet with $\lin=85$ and a pulse of 8 ps (a), 
The different fractional revivals are shown with vertical lines.}
\label{fig:frev}
\end{figure}

As it is well known, in addition to the revivals, there appear fractional revivals at times 
 \be{fractional}
 t = \frac{pT_r}{2(m^{2}-n^{2})},
 \ee
where the phase of the first cosine in \eq{ats-trunc} is a multiple of $\pi$. 
However, as the values of $n,m$ increase, the corresponding Gaussian weights decrease. 
Therefore, the higher the order of the fractional revival, the lower its intensity. An example is shown in \fig{fig:frev}.  Following the philosophy of keeping only the largest contribution, one can 
 directly see from \eq{ats-trunc}  how the partial revivals emerge by truncating the sum according to 
 the dominant contributions of the Gaussian weights.

\begin{figure}
\includegraphics[width=0.4\textwidth]{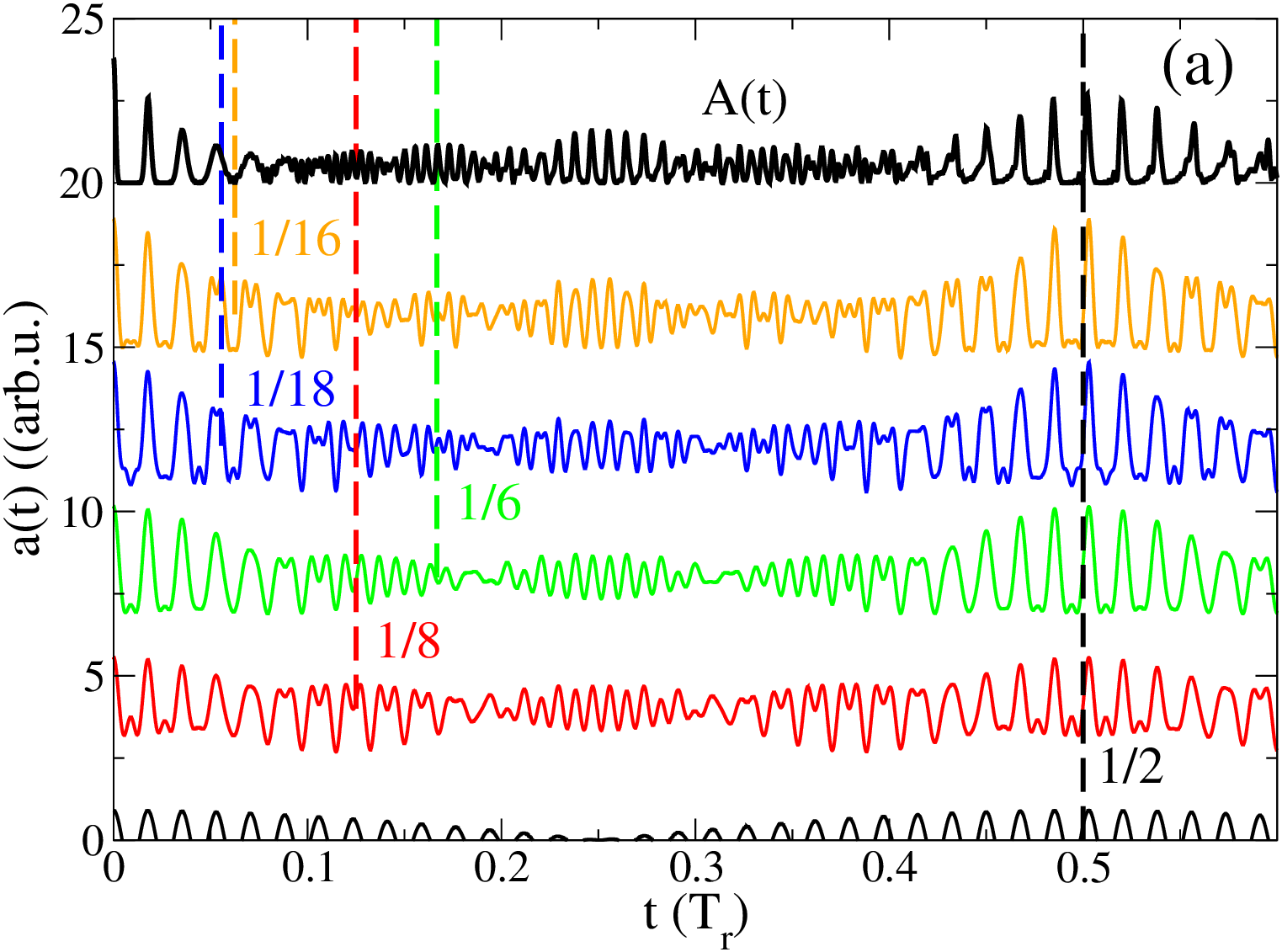} 
\includegraphics[width=0.4\textwidth]{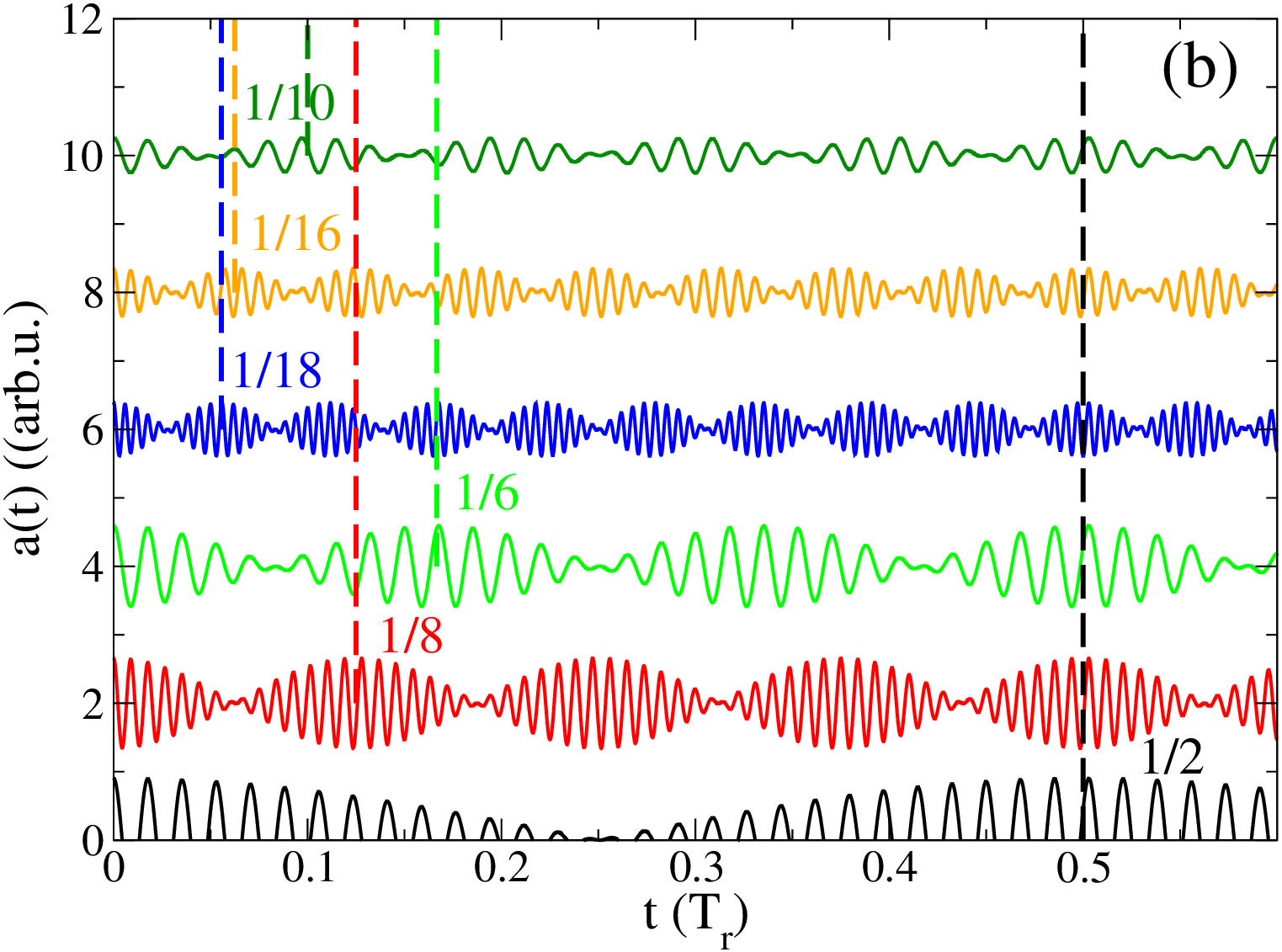} 
\caption{(Color online) $(a)$ Different truncations of the sum in \eq{ats-trunc},
for $Z=1$, $\lin=85$ and FWHM = 8 ps, including up to
$(n,m)=(1,0)$ (black), $(2,0)$ (red),
$(2,1)$ (green), $(3,0)$ (blue) and $(3,1)$ (orange). The exact analytical value A(t) (\eq{atorig}) is shown
for comparison at the top of the upper figure (thick black). The expected position of the fractional revivals is shown with vertical lines (see text). 
In $(b)$, the different components are shown separately (without summation) for the same cases plus for the pair $(3,2)$ (in dark green).}
\label{autocomp}
\end{figure}

From \tab{table1} we expect that  the largest contributions will arise from the fundamental pair
 $(n,m) = (1,0)$, which leads to the fractional revival at $T_{r}/2$. The second largest contribution comes from the  $(n,m) = (2,0)$  term, which has an exponent of $ n^{2}+m^{2} =4$. 
According to \eq{fractional} this contribution (with $p=1$) should give rise to a fractional revival at $T_{r}/8$. 
If the third largest term, the pair $(n,m) = (2,1)$ with an exponent of $5$ is included, a new partial revival will appear as expected at $T_{r}/6$, although with a very small weight.
This evolution can be seen in \fig{autocomp} for $Z=1$, $\lin=85$ and FWHM = 8 ps. In \fig{autocomp}a, the autocorrelation function emerges as more terms are added. The exact analytical value is shown in the top (thick black line).
The periodicities due to the different terms of the sum are shown in \fig{autocomp}b.

Overall, as can be seen from \eq{ats-trunc}, 
two factors determine the maximum order of fractional revivals resolved in the autocorrelation function:
the width of the pulse $\sigma_{T}$ and the energy difference between the states adjacent to $\lin$, 
as well as the inverse (discrete) classical revival time $T_{c}$ in the combination  $\sigma_{T}/T_{c}$:
The shorter 
 the pulse (small $\sigma_{T}$) and the smaller the energy differences (large $T_{c}$) the higher are these weights, and therefore more fractional revivals  will appear in these cases.

Higher values of $\lin$ with many states involved will create  higher orders of fractional revivals, 
as shown in \fig{fig:frev}(a) for a Rydberg wave packet with roughly 13 participating states (see table 1) with $Z=1$, $\lin=85$ and FWHM=8 ps.

\begin{table}
\caption{Exponents $m^{2}+n^{2}$ for the approximate weights in \eq{ats-trunc} 
of contributions $(m,n)$ giving rise to partial revivals $T_{r}/(m^{2}-n^{2})$, see \eq{fractional}.}
\label{table1}
\begin{ruledtabular}
\begin{tabular}{c|cc}
    pair & exponent & revival divisor \\
    $(m,n)$     & $m^{2}+n^{2}$   & ($m^{2}-n^{2}$)  \\\hline
    (1,0) & 1 & 1\\
        (2,0) & 4 & 4\\
      (2,1) & 5 & 3\\
           (3,0) & 9 & 9\\
               (3,1) & 10 & 8\\
                (3,2) & 13 & 5\\
\end{tabular}
\end{ruledtabular}
\end{table}
\section{Wavepacket dynamics in the low excitation domain}\label{sec:low}
We will now explicitly discuss the new regime of low quantum number wave packet dynamics.
As sketched in \fig{lowhigh2}b, for the case of a highly charged ion, i.e. with an effective atomic number $Z=8$ (\fig{lowhigh2}b), when the $\li{n}=14$ level is hit with a pulse of width
FWHM=2 fs, only $k\sim 4$ levels are excited and the relative energy difference between the levels is
large, thus one would expect periodicities obtained from a Taylor expansion to be less accurate.

The corresponding autocorrelation function (\eq{atorig}) is shown in \fig{autoRyd}b, both against $T_c$ (full lines),
and against $\bar{T}_c$ (dashed lines). The discrepancy is clear and becomes already seizable around the first revival, for which $\bar{T}_c$ describes very poorly the periodicity and $T_c$ remains almost exact. Also prominently visible are the maxima located at half-integer multiples 
of $T_c$ around $T_r/2$ and at integer multiples of $T_{c}$ close to the revival time $T_{r}$, as
explained in \sect{sec:signchange}.

\begin{figure}
\includegraphics[width=0.4\textwidth]{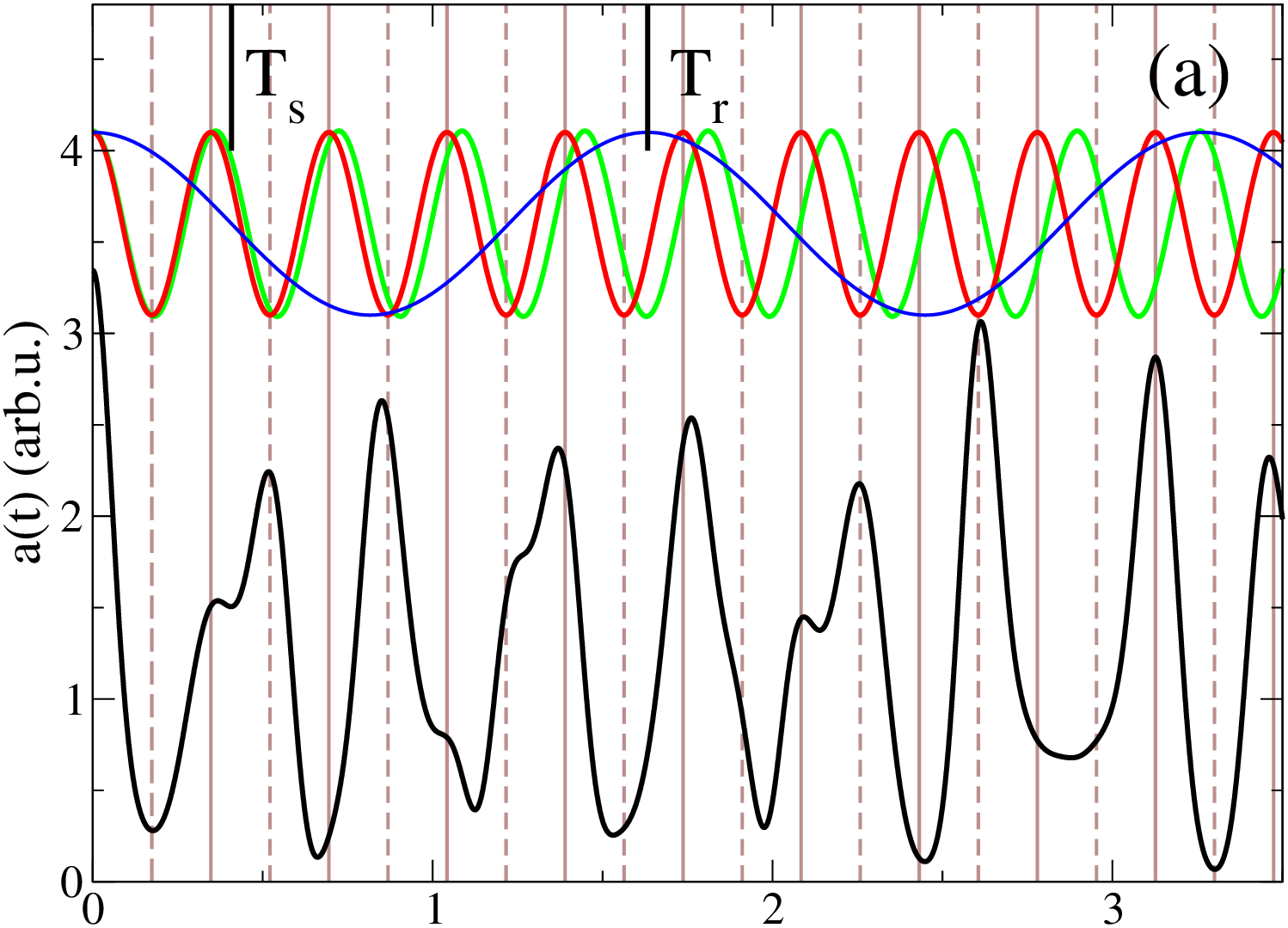} 
\includegraphics[width=0.4\textwidth]{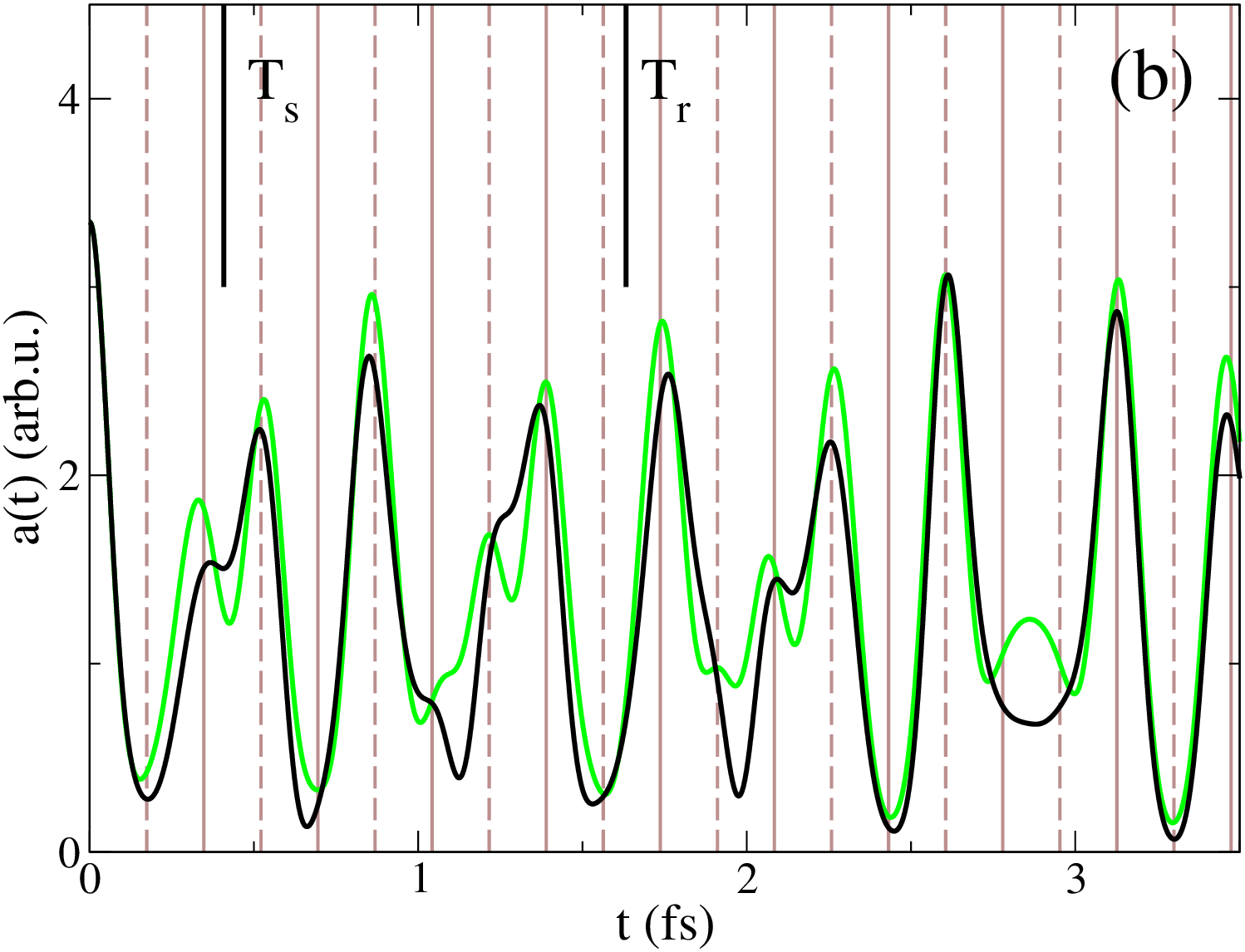} 
\caption{(Color online) Autocorrelation function for $Z=12$, $\li{n}=7$,  FWHM=0.1 fs.
(a) Black line: $a(t)$. Red line: $\cos(2\pi t/T_c)$. Green line: $\cos(2\pi t/\bar{T}_c)$. Blue line: $\cos(2\pi t/T_r)$.
The vertical lines are multiples (full lines) and half-multiples (dashed lines) of $T_c$.
(b) Same function compared with the case of varying dipole matrix elements $M_{1n}$, see text (green line).}
\label{autoZ12n7}
\end{figure}

The extreme case of a very low quantum number in the final state is shown in detail in \fig{autoZ12n7}. Here we consider a hydrogenic atom with
an effective $Z=12$ excited to $\lin=7$, with an ultrashort pulse of 0.1 fs with only 4 states participating,
a situation which one may also consider as a multi-quantum beat phenomenon.
The X-ray photon energy required for a $1s\rightarrow 7p$ transition in this system is
$\omega=1.92$ keV (we neglect the quantum defect).
While the sign change of the extrema at $T_{r}/4$ is very clearly visible, revivals can no longer be identified.
Also, the classical time $\bar{T}_{c}$ from a continuous energy derivative completely fails to predict the time scale (and therefore the position of the extrema) correctly after about 2 fs.

In \fig{autoZ12n7FFT}b we see that the particular value of the dipole elements $M_{in}$ has
a small influence on the autocorrelation spectrum in this regime. 
In this figure, the same autocorrelation function as in (a) is shown (black line) with all dipole elements  $M_{in}=1$, together
with the realistic calculation for $M_{in}=(2/n)^{3/2}$ (green line) \cite{Delone1994}. The details of $a(t)$ vary, but the position of the peaks remains overall the same.
\section{Non-resonant excitations and quantum defects}\label{sec:qd}
\begin{figure}
\includegraphics[width=0.4\textwidth]{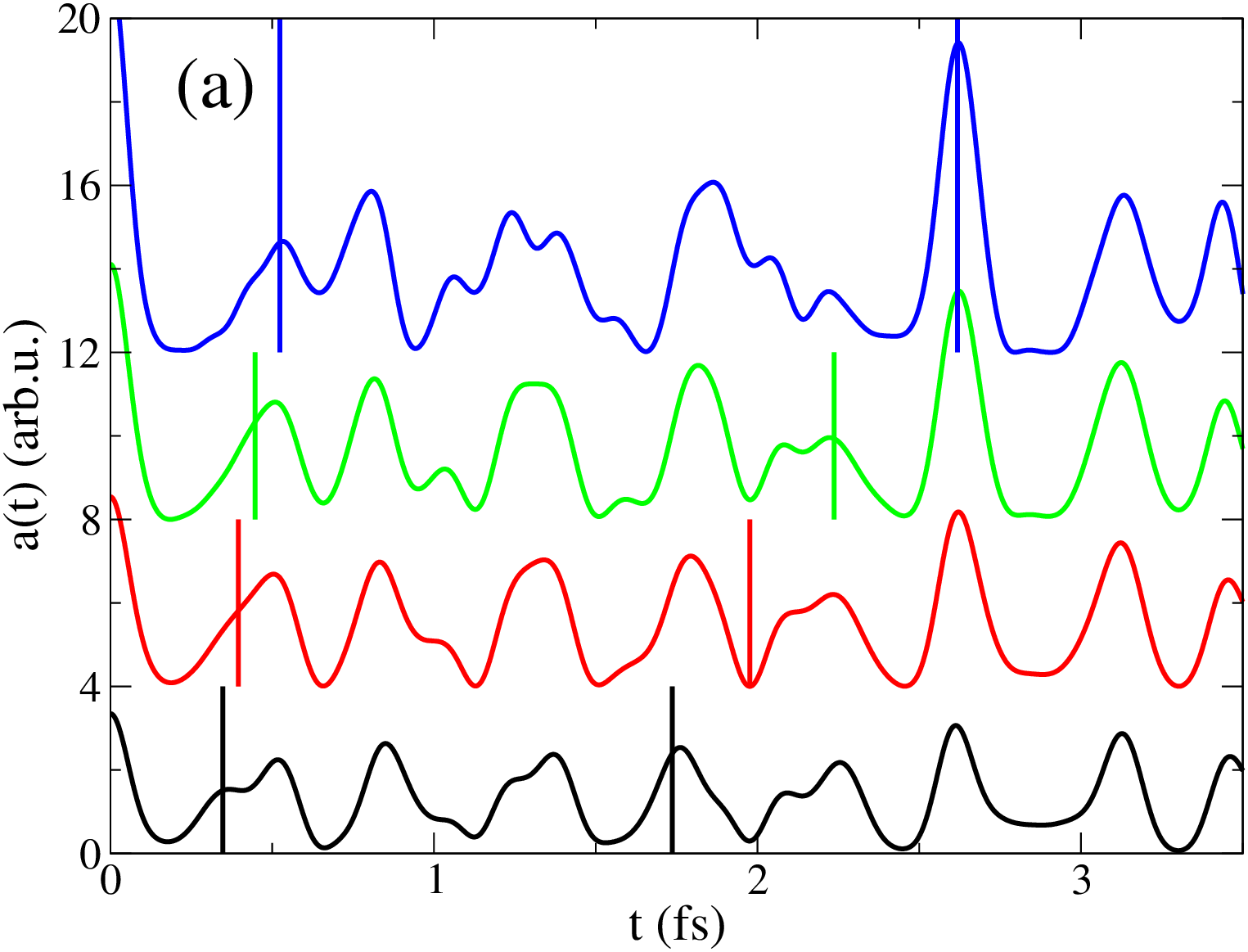} 
\includegraphics[width=0.4\textwidth]{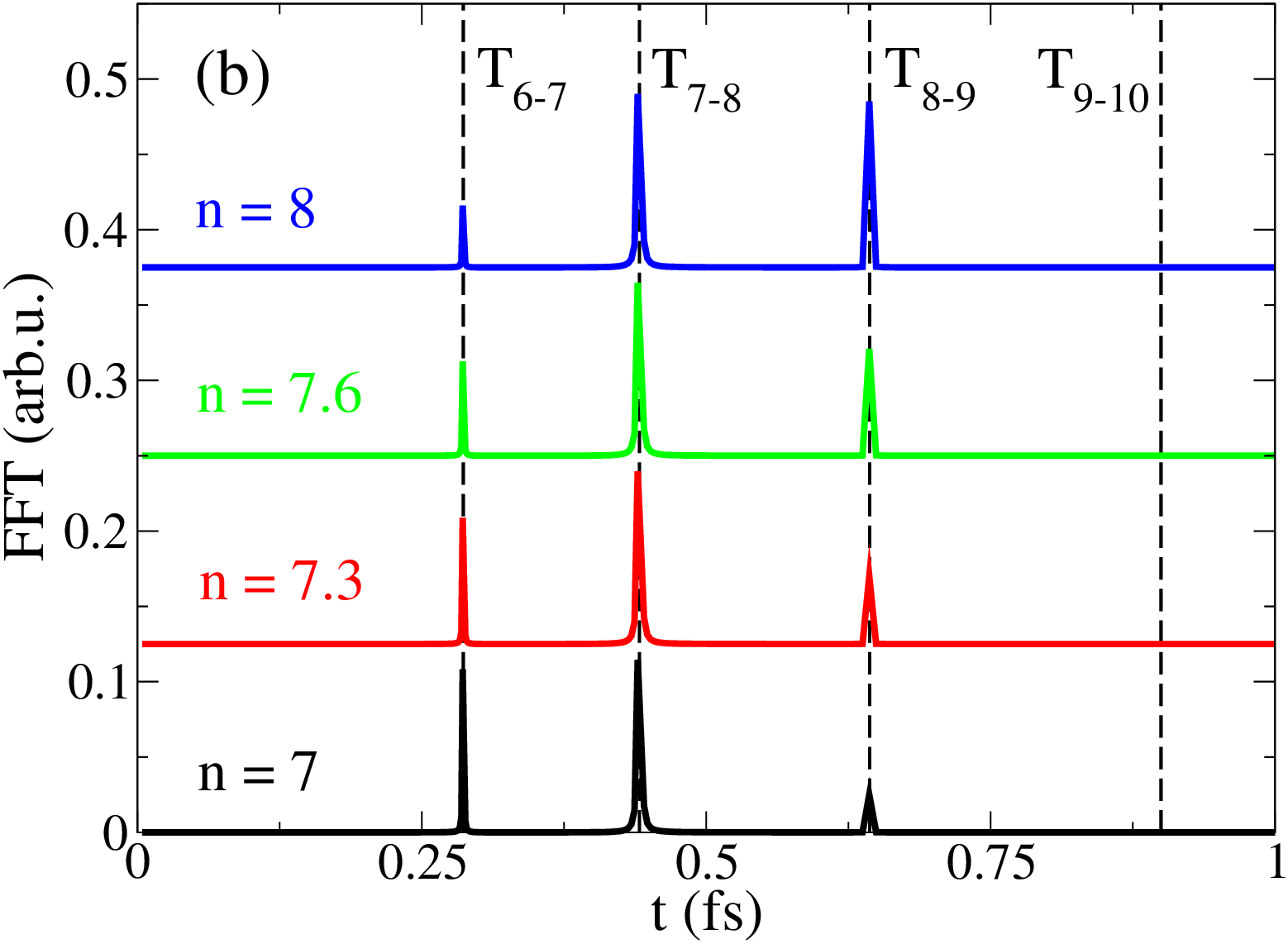} 
\caption{(Color online) (a) Same as in \fig{autoZ12n7}, but for four photon energies, corresponding to effective values
$\lin=7$, 7.3, 7.6 and 8. The values of $T_c$ and $5T_c$ are drawn in each case with vertical lines.
(b) Fourier transform of the cases in (a), using the two first terms in \eq{ats-trunc}. Abscissas are expressed in terms of time as $t=2\pi/\omega$. Vertical lines show the different periodicities, see text.}
\label{autoZ12n7FFT}
\end{figure}

In a general excitation process, the photon energy may not be resonant with any particular electronic transition. In this case 
\begin{equation}
E_1 + \Omega \rightarrow E_{\lin} + \epsilon.
\end{equation} 
We can identify this with an effective energy $E_{\lin+\Delta n}$,
where $0<\Delta n<1$. As $\Delta n$ grows, 
the central energy goes from $E_{\lin}$ to $E_{\lin+1}$, and the classical period evolves 
approximately
with the third power of $\lin+\Delta n$ (see \eq{TcRyd}). Notice that what really happens
is that while the pairs of eigenvalues involved in the different beatings remain the same, 
the relative weights in \eq{Gaussian} are changing, and so is $T_c$.
This is illustrated again for $Z=12$ and FWHM = 0.1 fs in \fig{autoZ12n7FFT}, for four different photon energies: 1.919 keV ($1s\rightarrow 7p$ transition), 1.922 keV ($\lin$=7.3), 1.925 keV ($\lin$=7.6) and 1.929 keV ($\lin$=8). 
The autocorrelation function and the Fourier transform are shown for the four cases in (a) and (b), respectively.
In the latter case, only two terms in \eq{ats-trunc} are used for simplicity, and abscissas are shown as a function of time. Vertical lines indicate the different periods, for $T_{a-b}=2\pi/(E_a-E_b)$. While $\lin=7$ and 8 show only two relevant contributions, more terms appear in the spectrum at non-integer values of $\lin$.

\begin{figure}
\centering
\includegraphics[width=0.35\textwidth]{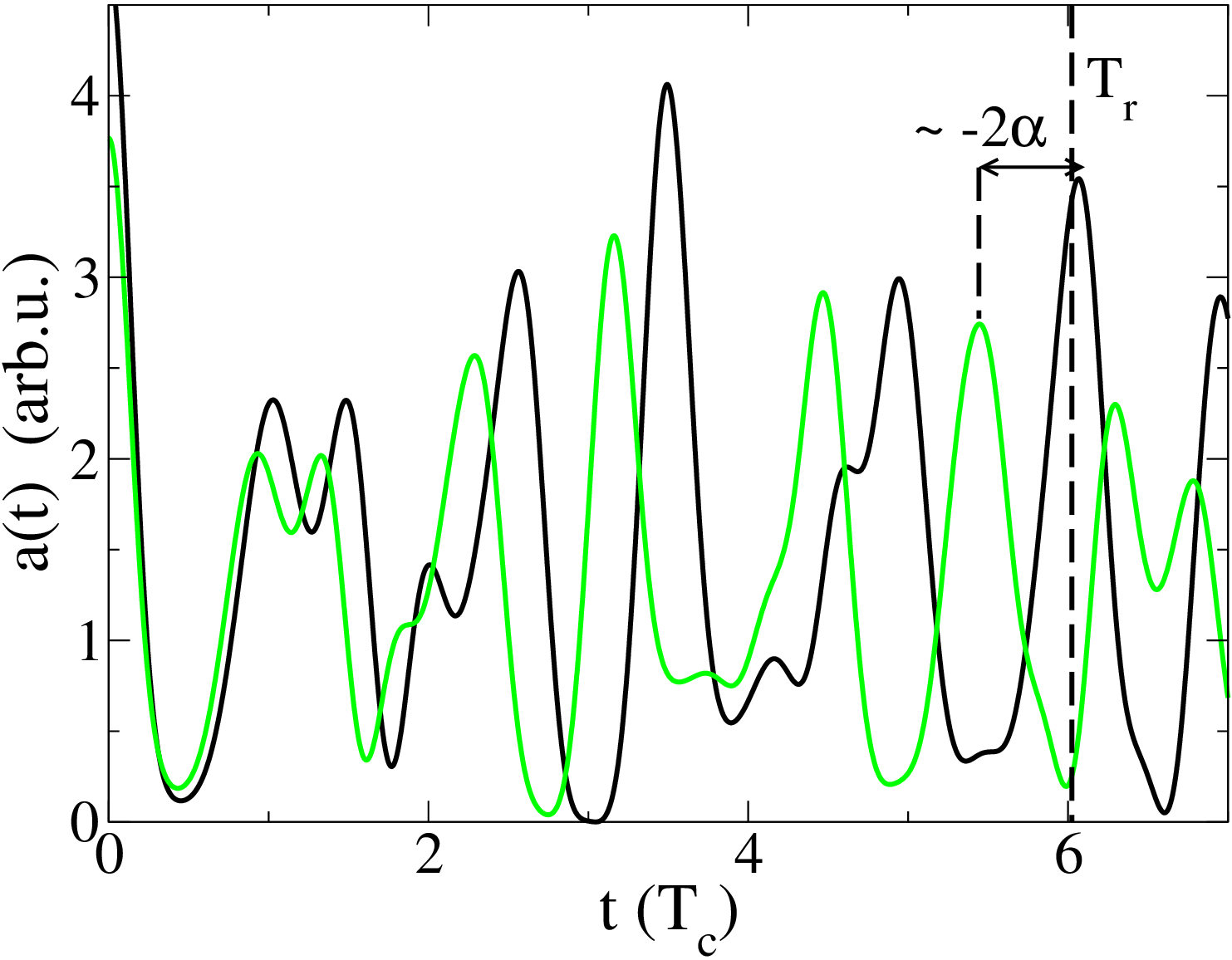} 
\caption{(Color online) Effect of the quantum defect on the autocorrelation function, for $Z=8$, $\lin=9$ and FWHM~=~0.4 fs, using  $\alpha=0$
(black) and $\alpha=0.3$ (green). Time is plotted in units of $T_c$, the classical period for the case with $\alpha=0$.
The shift in the peaks around $T_r$ is $\sim -2\alpha$.}
\label{atalpha}
\end{figure}

We have explored so far the hydrogenic case in which only one electron is bound to a nucleus of charge $Z$. 
However, if the system contains more than one electron, we might consider a nucleus of charge $Z+p$
and $p$ core electrons. 
The wave packet is then subject to an effective potential with a Coulombic tail of charge $Z$, whose eigenenergies are given by
\begin{equation}
E_n^\alpha=-\frac{Z^2}{2n_{\alpha}^2},
\end{equation}
where $n_{\alpha}=n-\alpha$. The quantum defect $\alpha$ depends on the angular momentum, but is almost independent on the quantum number $n$ for $n\gg 1$. 
The shift in the energies affects the periodicities of the system \cite{Bluhm1994,Wals1995}: 
With the period for the hydrogenic case ($\alpha=0$) and the multi-electron case ($\alpha\ne0$)
denoted by $T_c$ and  $T_c^{\alpha}$, respectively, the relative shift due to the quantum defect $\alpha$ is given by
\begin{eqnarray}
s_{\lin}(\alpha)&=&\frac{T_c^\alpha-T_c}{T_c} 
=\frac{(\li{n}_{\alpha}^2-1)^2}{(\li{n}^2-1)^2}\frac{\li{n}}{\li{n}_{\alpha}}-1 \nonumber\\
&=&\alpha\left[\frac{1+\li{n}_{\alpha}\li{n}(2-\alpha^2+3\alpha \li{n}-3\li{n}^2)}{\li{n}_{\alpha}(\li{n}^2-1)^2}\right].
\label{sal}
\end{eqnarray}
This shift accumulates in time and should assume the value 
\begin{eqnarray}
\frac{T_r}{T_c}s_{\lin}(\alpha)&=&\frac{2\li{n}^3}{3\li{n}^2-1}s_{\lin}(\alpha)
\sim-2\alpha
\label{salTr}
\end{eqnarray}
around $T_r$.
This prediction is valid even in the low $\lin$ regime, as is shown in \fig{atalpha} for $Z=8$, $\lin=9$ and a pulse length of 0.4 fs, using both\blue, $\alpha=0$ and 0.3.

\section{Conclusions}\label{sec:conc}
We have explored the dynamics of electron wave packets in the new regime of low atomic excitation which can be realized by short and intense laser pulses of high photon energy, e.g., from XFEL light sources. 
We have shown that the standard analysis of wave packet dynamics with characteristic periodicities carries over to the low excitation regime, but only when the periods are
calculated from differences of the two energy levels bracketing the main excitation level $\lin$ instead of using continuous derivatives.
Moreover, we have formulated the autocorrelation function in such a way that a new time scale
appears as an important structural element, namely a quarter of the revival time $T_{s}=T_{r}/4$, where
the sign of the extrema in the autocorrelation function flip sign. This is a prominent feature in low excitation wave packets and explains the ``shift'' in the maxima noticed previously in the high $\lin$ domain.
Hence, the analytical and numerical tools are now available to analyze  experiments on wave packet dynamics with intense short light pulses of high frequency.

JMR acknowledges support  in part by the National Science Foundation under Grant No. NSF PHY05-51164
during a visit of KITP at the University of Santa Barbara. 
PR acknowledges support in part from the MICINN projects FIS2010-15127 and JCI2009-03793, and the COST action CM0702.

\appendix\section{Regrouping the sum in the autocorrelation function} 
\subsection{Terms in the Taylor expansion}\label{app:terms}
The spectrum can be expanded around $E_{\lin}$:
\begin{equation}
E_{n+\lin}-E_{\lin} \equiv \delta_n = \delta^+_n + \delta^-_n
\end{equation}
The Taylor expansion is
\begin{eqnarray}
E_m - E_{n_0} &\sim&  \sum_{j=1}^{\infty} \frac{1}{j!}\frac{d^jE_m}{dm^j}\Big|_{m=n_0}(m-n_0)^j \nonumber \\
&=& \sum_{j=1}^{\infty} \frac{1}{(2j-1)!}\frac{d^{2j-1}E_m}{dm^{2j-1}}\Big|_{m=n_0}(m-n_0)^{2j-1}\nonumber \\
&&+ \sum_{j=1}^{\infty} \frac{1}{(2j)!}\frac{d^{2j}E_m}{dm^{2j}}\Big|_{m=n_0}(m-n_0)^{2j}
\end{eqnarray}
At $m=n+n_0$ it is
\begin{eqnarray}
E_{n+n_0} - E_{n_0} &\sim&  
\sum_{j=1}^{\infty} \frac{1}{(2j-1)!}\frac{d^{2j-1}E_{n+n_0}}{dn^{2j-1}}\Big|_{n=0}n^{2j-1}\nonumber \\
&&+ \sum_{j=1}^{\infty} \frac{1}{(2j)!}\frac{d^{2j}E_{n+n_0}}{dn^{2j}}\Big|_{n=0}n^{2j},
\end{eqnarray}
so if we define 
\begin{equation}
\frac{1}{\bar{T}_j} \equiv \frac{1}{2\pi j!}\frac{d^jE_{n+n_0}}{dn^j}\Big|_{n=0},  
\end{equation}
then 
\begin{eqnarray}
E_{n+n_0} - E_{n_0} &\sim&  2\pi\sum_{j=1}^{\infty}\frac{n^{2j-1}}{\bar{T}_{2j-1}}
+2\pi\sum_{j=1}^{\infty}\frac{n^{2j}}{\bar{T}_{2j}} \nonumber \\
&\equiv & \delta^-_n + \delta^+_n.
\end{eqnarray}

\subsection{Final expression for $a(t)$}\label{app:final}
An expansion of the cosines in \eq{ats1} brings
\begin{eqnarray}
a(t)/2  &=& \sum_{0\leq n<m}
W_{mn}\Big[\cos [(\delta^+_{m}-\delta^+_{n})t]\cos [(\delta^-_{m}-\delta^-_{n})t] \nonumber \\
&&-\sin [(\delta^+_{m}-\delta^+_{n})t]\sin [(\delta^-_{m}-\delta^-_{n})t]\Big] \nonumber \\
&& +W_{-m-n}\Big[\cos[ (\delta^+_{-m}-\delta^+_{-n})t]\cos[ (\delta^-_{-m}-\delta^-_{-n})t] \nonumber \\
&&-\sin[ (\delta^+_{-m}-\delta^+_{-n})t]\sin[ (\delta^-_{-m}-\delta^-_{-n})t] \Big]. 
\label{at2exp}
\end{eqnarray}
Due to the symmetric properties of the energy differences, it is $\delta^+_{-n}=\delta^+_n$ and
$\delta^-_{-n}=-\delta^-_n$, so
\begin{eqnarray}
\delta^+_{-m}-\delta^+_{-n} &=& \delta^+_{m}-\delta^+_{n}, \\
\delta^-_{-m}-\delta^-_{-n} &=& -(\delta^-_{m}-\delta^-_{n}).
\end{eqnarray}
The coefficients are
\begin{equation}
W_{nm} = e^{-\left[ (\delta_m^2 + \delta_n^2)/\sigma^2 \right]}  
= e^{-w_{mn}/\sigma^2}e^{-2v_{mn}/\sigma^2},
\end{equation}
where $w_{mn}$ is defined in \eq{ats2a}, and
\begin{equation}
v_{mn} \equiv \delta^+_m\delta^-_m + \delta^+_n\delta^-_n.
\end{equation}
Again due to symmetry reasons, $w_{-m-n}=w_{mn}$ and $v_{-m-n}=-v_{mn}$, and thus
\begin{equation}
W_{-n-m} = e^{-w_{mn}/\sigma^2}e^{2v_{mn}/\sigma^2}.
\end{equation}
With this, \eq{at2exp} becomes
\begin{eqnarray}
a(t)/2  &=& \sum_{0\leq n<m}
W_{mn}\Big[\cos [(\delta^+_{m}-\delta^+_{n})t]\cos [(\delta^-_{m}-\delta^-_{n})t] \nonumber \\
&&-\sin [(\delta^+_{m}-\delta^+_{n})t]\sin [(\delta^-_{m}-\delta^-_{n})t]\Big] \nonumber \\
&& +W_{-m-n}\Big[\cos[ (\delta^+_m-\delta^+_n)t]\cos[ (\delta^-_m-\delta^-_n)t]\nonumber\\
&&+\sin[ (\delta^+_m-\delta^+_n)t]\sin[ (\delta^-_m-\delta^-_n)t] \Big]. 
\label{at2expC}
\end{eqnarray}
The terms with cosines gather in a term with coefficient
\begin{eqnarray}
W_{mn}+W_{-m-n}&=&e^{-w_{mn}/\sigma^2}\left( e^{-2v_{mn}/\sigma^2}+e^{2v_{mn}/\sigma^2} \right) \nonumber\\
&=&2e^{-w_{mn}/\sigma^2}\cosh{\left[\frac{2v_{mn}}{\sigma^2}\right]},
\end{eqnarray}
while the terms with sines gather with a coefficient
\begin{eqnarray}
-W_{mn}+W_{-m-n}&=&e^{-w_{mn}/\sigma^2}\left( -e^{-2v_{mn}/\sigma^2}+e^{2v_{mn}/\sigma^2} \right) \nonumber\\
&=&2e^{-w_{mn}/\sigma^2}\sinh{\left[\frac{2v_{mn}}{\sigma^2}\right]}.
\end{eqnarray}
Therefore,
\begin{eqnarray}
(W_{mn}+W_{-m-n})e^{w_{mn}/\sigma^2} = 2\sum_{j=0}^{\infty}\frac{(2v_{mn})^{2j}}{(2j)!\sigma^{4j}},
\end{eqnarray}
and
\begin{eqnarray}
(-W_{mn}+W_{-m-n})e^{w_{mn}/\sigma^2} = 2\sum_{j=0}^{\infty}\frac{(2v_{mn})^{2j+1}}{(2j+1)!\sigma^{4j+2}},
\end{eqnarray}
which are respectively $2C_{mn}$ and $2S_{mn}$, as defined in \eq{ats2a}.
This way we obtain \eq{ats2}:
\begin{eqnarray}
a(t)/4  &=& \sum_{0\leq n<m}e^{-\frac{w_{mn}}{\sigma^2}}
\Big[ C_{mn}
\cos [(\delta^+_{m}-\delta^+_{n})t]\cos [(\delta^-_{m}-\delta^-_{n})t] \nonumber \\
&&+S_{mn}
\sin [(\delta^+_{m}-\delta^+_{n})t]\sin [(\delta^-_{m}-\delta^-_{n})t]\Big].
\label{ats2-app}
\end{eqnarray}

\subsection{Lowest order of the expansion}\label{app:lowest}
If we take the lowest order ($j=1$) in the definition of $\delta^+_m$ and $\delta^-_m$ 
at \eq{truncation}, then 
\begin{eqnarray}
\delta^+_m - \delta^+_n & =& \frac{2\pi}{\bar{T}_2}(m^2-n^2),\nonumber\\
\delta^-_m - \delta^-_n & =& \frac{2\pi}{\bar{T}_1}(m-n),
\end{eqnarray}
and
\begin{eqnarray}
w_{mn} &=& (\delta^{+}_{m})^{2} +(\delta^{-}_{m})^{2}+(\delta^{+}_{n})^{2}+(\delta^{-}_{n})^{2}\nonumber\\
&=& 4\pi^2\left(\frac{m^4}{\bar{T}_2^2} +\frac{m^2}{\bar{T}_1^2} +\frac{n^4}{\bar{T}_2^2}+\frac{n^2}{\bar{T}_1^2} \right),
\end{eqnarray}
which at second order in the derivatives is
\begin{eqnarray}
w_{mn}
&\sim& \frac{4\pi^2}{\bar{T}_1^2}(m^2+n^2).
\end{eqnarray}
The lowest order of the sums in $C_{mn}$ and $S_{mn}$ is $j=0$, so
\begin{eqnarray}
C_{mn} &=& 1, \nonumber\\
S_{mn} &=& \frac{2v_{mn}}{\sigma^2}.
\end{eqnarray}
Since $S_{mn}=0$ at second order in the derivatives, and using $\sigma_T=1/\sigma$, we obtain \eq{ats-trunc},
\begin{align}\label{ats-trunc-app}
 a_{2}(t)/4 \sim \sum_{0\leq n<m}&e^{-\frac{4\pi^{2}\sigma_{T}^{2}}{\bar{T}_{1}^{2}}(m^{2}+n^{2})}
 \cos\left[2\pi(m^{2}-n^{2})\frac{t}{\bar{T}_{2}}\right] \nonumber \\
&\times\cos\left[2\pi(m-n)\frac{t}{\bar{T}_{1}}\right]\,.
\end{align}

\end{document}